\documentstyle[12pt]{article}

\newcommand{\gsim}{\stackrel{>}{\sim}}  

\def\psibar{{\bar \psi}}

\begin{document}

\begin{titlepage}
\begin{flushright}
\hfil                 UCSD/PTH 92-18\\
\hfil June 1992 \\
\end{flushright}
\begin{center}

\vspace{1cm}
{\LARGE\bf Chiral Fermions and Anomalies \\
\vspace{0.5cm}
on a Finite Lattice}

\vspace{1.5cm}
{\large Karl Jansen}\\
\vspace{1cm}
University of California at San Diego\\
Department of Physics-0319 \\
La Jolla, CA~92093-0319 \\
USA  \\
\vspace{0.2cm}

\vspace{1.5cm}
\end{center}

\abstract{
\noindent Recently Kaplan proposed a new method to
simulate chiral fermions
on the lattice by introducing a space dependent mass term that looks
like a domain wall.
He showed that if one
starts with an odd dimensional lattice theory
the lower (even) dimensional world on the domain wall
will have a chiral zero mode and the corresponding anomaly.
I test this proposal by computing the
gauge and Goldstone-Wilczek currents
and their derivatives on a
3-dimensional finite lattice. By determining the spectrum of the finite
lattice Hamiltonian I
demonstrate that the theory one obtains is chiral below a critical
momentum.
Furthermore I show that one can see
the anomaly on the finite lattice and that
in the case of the 3-4-5 model the anomalies in the gauge currents
cancel as in the continuum.
}
\vfill
\end{titlepage}

\section{Introduction}

The Standard Model and several of its extensions are chiral theories
and consequently chiral
gauge theories play an important role in our theoretical concepts.
In principle one would like to understand these theories not
only perturbatively but also at strong coupling. However, it has turned
out that a non-perturbative regularization of chiral gauge theories runs
into problems. If one wants
to use the lattice as a regulator one faces
the appearance of extra fermion species,
the doubler fermions. Although it is a general belief
that the lattice can be used for regulating every quantum field theory,
it seems to fail in the case of chiral gauge theories \cite{smit}.
This failure is summarized in the Nielsen-Ninomiya theorem \cite{nn} which
leaves us with the choice of either having
the unwanted doubler fermions on the
lattice or no doubler modes but also the lack of chiral symmetry.
Of course,
the problem with the lattice as a regulator for chiral
gauge theories raises the question, of whether this failure can be
attributed only to the lattice or if it is a generic problem also with
other non-perturbative regulators for chiral gauge theories.

The last years have seen numerous attempts to latticize the Standard
Model \cite{smit}. Some of them like the Smit-Swift proposal have
been carefully analyzed and tested, both in numerical and analytical
computations. However, it is generally accepted by now
\cite{goltermann,petcher} that the
result of these investigations is negative. So far a successful method
to put chiral fermions on the lattice is missing.

Here I want to start an investigation of a recent proposal by
D.~Kaplan \cite{kaplan}.
He started with an odd-dimensional {\em vector} gauge theory,
avoiding in this way problems with the Nielson-Ninomiya theorem. In this
odd dimensional theory
the fermion mass is taken to depend on one of the coordinates with the
structure of a domain wall. Following Kaplan,
on this domain wall --which is a
lower (even) dimensional world-- one should find a {\em chiral} fermion, the
correct anomaly structure and the decoupling of the doubler modes.
Therefore for a (lattice) person who lives on the domain wall and does not
know about the extra dimension with the defect, it would seem as if the
Nielsen-Ninomiya theorem is violated
whereas a person living in the higher dimensional world would interprete
this result as a consequence of the existence of the domain wall.

In this letter I provide a first test of Kaplan's proposal.
By using numerical methods I calculate the spectrum
and --in the presence of external gauge fields-- the currents
on the finite lattice. In
particular I demonstrate the existence of the
anomaly on a finite lattice.

\pagebreak
\section{The model}

To be specific I will restrict myself to a 3-dimensional U(1) gauge theory
on a lattice $L^2L_s$.
The action is given by
\begin{eqnarray}
S & = & \frac{1}{2}
     \sum_{z,\mu} \left[ \psibar_z \gamma_\mu U_{z,\mu} \psi_{z+\mu}
                 -\psibar_{z+\mu}\gamma_\mu U_{z,\mu}^* \psi_z
\right] +m_0(s) \sum_z \psibar_z\psi_z \nonumber \\
 & + & w\sum_{z,\mu} \left[ -2\psibar_z\psi_z
+\psibar_xU_{z,\mu}\psi_{z+\mu} + \psibar_{z+\mu}U_{z,\mu}^* \psi
               \right] \;\; .
\label{eqnarray:action}
\end{eqnarray}
\noindent Here a lattice point is denoted by $z=(t,x,s)$
and $\mu$ is a unit vector pointing in one of the three directions.
The lattice spacing $a$ is set to one throughout the paper. The
fermion fields are 2-component complex spinors and the gauge fields
$U_{z,\mu}\in U(1)$.
They are related to the vector potential by  $U_{z,\mu} = e^{iqA_\mu(z)}$
with $q$ the electric charge.
I leave out the gauge field self interaction (i.e. the
plaquette term) as I am only interested in external gauge fields at
this point.
The gamma matrices in euclidean space are given by the Pauli-matrices.
Note that in odd dimensions there is no analogue of $\gamma_5$
so that the theory is automatically vector-like.
The action (\ref{eqnarray:action}) has a mass term depending only on one
of the lattice directions, $s$:

\begin{equation}
m_0(s) \equiv \sinh(\mu_0)\theta(s) ;\;\;
\theta(s) = \left\{ \begin{array}{lll}
                    -1 & 2\le s \le \frac{L_s}{2} \\
                    +1 & \frac{L_s}{2}+2 \le s \le L_s \\
                     0 & s=1, \frac{L_s}{2}+1
                     \end{array} \right. \;\; .
\label{eq:m0}
\end{equation}

\noindent This mass has the form of a kink-antikink pair
and generates two domain
walls, one located at $s=1$ and the other at $s=\frac{L_s}{2}+1$.
The existence of the antikink is necessitated by periodic boundary
conditions.
When one increases the lattice size the
distance between these domain walls which is $L_s/2$, increases and for
$L_s =\infty$ we are left with only one domain wall.

To get some insight into the spectrum of the theory,
let me turn off the gauge fields for a moment and set the Wilson
coupling $w$ to be zero (For a discussion of zero modes on domain walls
see also \cite{jarebbi,fradkin}).
Then the lattice Hamiltonian of the problem is

\begin{equation}
H=-\sigma_1\left[\sigma_2\partial_x + \sigma_3\partial_s +m_0(s)\right]\; ,
\label{eq:hamiltonian1}
\end{equation}
\noindent where $\partial$ denotes the lattice derivative $\partial_z =
\frac{1}{2}\left[\delta_{z,z+\mu} - \delta_{z,z-\mu}\right] $.

It is easy to see that
\begin{equation}
\psi^+ = e^{ikx}e^{+\mu_0|s-(\frac{L_s}{2}+1)|}\left( \begin{array}{l}
                                   1 \\
                                   0
                                   \end{array} \right)
\label{eq:psiplus}
\end{equation}
is an (unnormalized) eigenstate
of the Hamiltonian, which is a massless fermion with
chirality $\sigma_3\psi^+ = + \psi^+$. This solution is centered on the
domain wall at $s=1$ and falls off exponentially as one goes away from
the wall.
There is a similar solution $\psi^-$
\begin{equation}
\psi^- = e^{-ikx}e^{-\mu_0|s-(\frac{L}{2}+1)|}\left( \begin{array}{l}
                                     0 \\
                                     1
                                     \end{array} \right)
\label{psiminus}
\end{equation}
which lives on the domain wall at $s=\frac{L_s}{2}+1$ with negative
chirality.
Without the Wilson-term one has also the doubler modes at the corners of
the Brillouin zones.
They remain massless and appear with the opposite chirality.
Therefore the fermion and the doubler modes pair up, the resulting
theory is not chiral and so far we have gained nothing.

To decouple the doublers Kaplan added the usual Wilson-term,
which in this case is not problematic as we started with a
vectorlike theory.
Introducing the Wilson-term leads to
\begin{equation}
H  =-\sigma_1\left[\sigma_2\partial_x + \sigma_3\partial_s
+m_0(s)+w(\Delta_x +\Delta_y) \right] \;\; .
\label{eq:hamiltonian}
\end{equation}
Here $\Delta$ denotes the second derivative on the lattice.
The Hamiltonian (\ref{eq:hamiltonian}) can be reduced to only depend on
$s$ by imposing plane wave solutions in the $x$-direction. One obtains
\begin{equation}
H  =-\sigma_1\left[\sigma_2\sin(k) + \sigma_3\partial_s
+m_0(s)+2w(\cos(k)-1) +w\Delta_s) \right] \;\; .
\label{eq:hamiltonian2}
\end{equation}

In \cite{kaplan} it was shown that for the infinite system
which has a single domain wall, the Wilson term gives a large mass to the
would-be chiral modes when their momenta exceed a critical value $k_c
=O(1)$. Because the aim at the end is a finite lattice simulation
one wants to keep the same
scenario at finite volume.
Of course, on the finite lattice one has
two domain walls and
expects two chiral modes, one located at the domain
wall at $s=1$ with positive chirality and the other located at $L_s/2+1$ with
negative chirality.
The effect of the Wilson term is to delocalize and
pair up high momentum bound states of opposite chirality on the two
doamin walls. For low momentum, the states remain chiral and massless
and their overlap is exponentially small $\approx e^{-\mu_0L_s/2}$.
If we look at only one domain wall we are left with
only a single chiral mode in the low energy regime.

To see whether this scenario is true I calculated the
eigenvalues and eigenfunctions of the Hamiltonian
(\ref{eq:hamiltonian2}) numerically on a system size of $L=L_s=100$.
The lattice momenta $k$ are given by
$k=\sin((\pi(n +\frac{1}{2}))/L)$
where $n=0,...,L-1$. These momenta
correspond to
antiperiodic boundary conditions in the $x$-direction
as will be used for the calculation of
the anomaly in the next section.
Let me discuss the results of the numerical
evaluation of the eigenvalues and the eigenfunctions.
The eigenvalues occur in $\pm\lambda(k)$ pairs
corresponding to a particle and an anti-particle solution.
I plot the lowest momentum, $k=\sin(\pi/2L)$, eigenfunctions
which correspond to the lowest eigenvalues $\pm\lambda_0$
in fig.1. The figure shows that
the $+\lambda_0$ state is located at $s=1$, while the $-\lambda_0$
state is localized at $s=\frac{L_s}{2}+1$.
They are well separated in agreement with
the expected overlap of the wavefunctions
which is $\approx e^{-50}$ ($\mu_0$ was chosen to be $0.81$).

In Fig.2a I show the lowest two eigenvalues
$\lambda_0(k)$ and $\lambda_1(k)$ of the Hamiltonian
(\ref{eq:hamiltonian2}).
For $k < k_c \approx 0.9$, $\lambda_0$ exhibits the dispersion relation
of a massless fermion while $\lambda_1$ corresponds to a state with a
value of the mass at the cut-off and the system has a mass gap.
Note that $k_c$ is already at the order of the cut-off.
Increasing the momenta $k$ above $k_c$ the eigenvalues degenerate and
the energies are above the cut-off.
The doubler
modes at the corner of the Brillouin zone are decoupled and there is
only one zero mode in the spectrum.

Fig.2b shows that this zero mode is a chiral fermion. I plot the ratio
\begin{equation}
R= \frac{<\psibar \psi >}{<\psibar \sigma_1 \psi >}.
\label{eq:ratio}
\end{equation}
\noindent This ratio measures the ``violation of chirality''.
It is zero if $\psi$ is chiral.  $R$ being non-zero
indicates that the state is no longer a chiral eigenstate.
The figure clearly
demonstrates that for small momenta the wavefunction on the lattice is
chiral. For large values of $k \gsim 0.9$ the chirality is lost, the
large momentum (at the order of the cut-off) modes pair up.

I conclude that if I am restricted to the domain wall at $s=1$ I find a
chiral fermion at low energies on the finite lattice. As a consequence I
should see the corresponding anomaly. That this is indeed the case will
be shown in the next section.

\section{The anomaly}

\subsection{In the continuum}

It is well known that the existence of a zero mode on a mass defect
produces an anomaly \cite{gw,caha}. This anomaly should appear
in the 2-dimensional world on the domain wall and is given by
\cite{anomaly}
\begin{equation}
\partial_i j_i(z) = \pm q^2 \frac{E(t)}{2\pi}\;\; .
\label{eq:anomaly1}
\end{equation}
\noindent Here $j_i$ is the current $j_i=<\psibar \gamma_i \psi >$
, ($i=t,x$),
$q$ is the charge of the fermion, $E(t)$ is an applied
external $E$-field and the $\pm$ signs stand for the chirality of the zero
mode on the domain wall.
(The factor of $i$ appearing in the Euclidean anomaly has been absorbed
into the definition of the current $j$.)
The existence of this anomaly may seem surprising because we
started from a
3-dimensional vector gauge theory and no anomaly should be
present. There must be
another current wih nonzero divergence that
cancels the zero mode anomaly (\ref{eq:anomaly1}).

This extra contribution is found in the Goldstone-Wilczek current \cite{gw}.
It is well known that charge can flow due to the adiabatic change of
external fields. In the present example this
effect creates the so-called Goldstone-Wilczek current
\cite{gw,caha}, along the $s$-direction,
\begin{equation}
j_s^{GW}(z) = -\frac{q^2}{2\pi} \frac{m(s)}{|m(s)|}E(t)\;\; .
\label{eq:gwcurrent}
\end{equation}
This current is perpendicular to the
applied $E$-field (resembling in this respect a Hall current) and to the
domain wall. As the mass $m_0(s)$ is different on the two sides of the
domain wall, the divergence of this current is not zero and
exactly opposite to the divergence of the current
(\ref{eq:anomaly1}). The 3-dimensional vector theory is anomaly free.
The cancellation of the currents (\ref{eq:anomaly1}) and (\ref{eq:gwcurrent})
is an example of the general relation of anomalies in odd dimensions to
the ones in even dimensions as discussed in \cite{caha}.

If the picture described above is correct, applying an external
$E$-field one has to see the anomaly
(\ref{eq:anomaly1})
even on a finite lattice.
Moreover, one should have no anomaly if one
starts with an anomaly free theory like the 3-4-5 model as discussed in
\cite{kaplan}.
In this model one chooses as the mass term
\begin{equation}
m_0(s)\psibar_z\psi_z \rightarrow
m_0(s)\left[(\psibar_z\psi_z)_3 + (\psibar_z\psi_z)_4 -
(\psibar_z\psi_z)_5\right] \;\; .
\label{eq:345}
\end{equation}

\noindent The indices here mean that the fermions have charge $q=3,4,5$,
respectively. Because of the
minus sign of the $q=5$ fermion mass
the $q=5$ fermion and the $q=3,4$ fermions have opposite chirality.
Furthermore, because the anomaly is proportional to the charge squared, the
sum of the individual divergencies should vanish.
This would not only test that one sees the anomaly on the lattice
but also that it is proportional to $q^2$ and has therefore the right
amplitude.

\subsection{On the lattice}

To test the above picture I have calculated the gauge current and its
derivative numerically.
The current on the finite lattice is given by
\begin{equation}
j_z^\mu = \frac{1}{2}\left[\psibar_z \gamma_\mu U_{z,\mu} \psi_{z+\mu}
      + \psibar_{z+\mu}\gamma_\mu U_{z,\mu}^* \psi_z \right]+
  w \left[\psibar_zU_{z,\mu}\psi_{z+\mu}
   - \psibar_{z+\mu}U_{z,\mu}^* \psi \right]
\label{eq:latticecurrent}
\end{equation}

\noindent where $\mu=1,2,3$. Note that I leave out the $i$ in front of
the current, because it will drop out in the anomaly equation
(\ref{eq:anomaly1}).
The total divergence of this current is locally
conserved $\partial_\mu j^\mu_z =0$
because the theory (\ref{eqnarray:action}) is vectorlike.
But if we pretend to live in the
2-dimensional world on one of the domain walls,
where we have a chiral fermion,
we would see the anomaly (\ref{eq:anomaly1}) realized.

The current (\ref{eq:latticecurrent}) can be calculated numerically on a
finite lattice from the inverse fermion matrix.
I used the conjugate gradient method to invert the matrix in the
presence of the external gauge fields.
Note that this step is a numerical computation and that {\em no}
simulation is involved.
I have chosen antiperiodic
boundary conditions in the $x$ direction to avoid problems with zero
modes that would render the matrix not invertible. To implement an
external $E$-field along the domain wall, the gauge fields were chosen
to be
\begin{equation}
U_{x,\mu=2} = \exp\{-iq\left[\frac{L}{2\pi}E_0\cos(\frac{2\pi}{L}(t-1))
  \right]\}
\label{eq:efield}
\end{equation}
The $U$'s in the other directions were set to 1.

As is shown in fig.1, the wavefunctions on the lattice have a finite
width. The $1+1$-dimensional world is not localized at only
$s=1$ but extends over several $s$-slices.
The charge is built up in this finite region in $s$ and
the anomaly equation
(\ref{eq:anomaly1}) has therefore to be modified on the lattice. Let us
define
\begin{equation}
<\partial_ij_i> \equiv \sum_{s\in \Lambda_\psi}
\partial_ij_i(t,x,s)\;\;, i=(t,x).
\label{eq:latticedef}
\end{equation}
\noindent Here the sum in s is taken over the support $\Lambda_\psi$ of the
wavefunction (see fig.1)
--the ``width'' of the $1+1$-dimensional world.
By varying the number of $s$-slices one can
determine how many slices one has to take in the summation so that the
result for $<\partial_ij_i>$ does not change.
The anomaly equation on the lattice reads now
\begin{equation}
<\partial_ij_i> = \pm\frac{q^2}{2\pi}E_{eff}(t)
\label{eq:latticeanomaly}
\end{equation}
\noindent where
the effective electric field $E_{eff}$ for small $E_0$ is given by
\begin{equation}
E_{eff}=\frac{\sin(\frac{2\pi}{L})}{\frac{2\pi}{L}}E_0\sin(t-1)\; .
\end{equation}

One can perform
several checks on the program.
First, using the wave function (\ref{eq:psiplus})
as an input and
acting on it with the fermion matrix it should reproduce the zero
eigenvalue if the gauge fields are switched off. This is an
excellent test to see whether the correct matrix is in the program.
Second, because the 3-dimensional gauge current is conserved,
the total divergence of the current should cancel {\em locally} in the
presence of the external E-field. Also this I
could see clearly by calculating the {\em local} currents. It turned out
that $\partial_xj_x$ was exactly zero (as one would expect) and that the
cancellation only occured with $\partial_tj_t$
(the zero mode current (\ref{eq:anomaly1}))
and $\partial_sj_s$ (the Goldstone-Wilczek
current (\ref{eq:gwcurrent})) at the same
lattice point, i.e.
$\partial_xj_x =0 \;\;,\partial_tj_t = -\partial_sj_s$.

I have chosen a $16^3$ lattice with $\mu_0 =0.81$ (see (\ref{eq:m0}))
and the strength of the $E$-field $E_0=0.001$ \footnote{The choice of
$E_0$ was motivated to stay in the low energy regime of the dispersion
relation (see fig.2a).}.
The outcome of the numerical evaluation of the current in the presence
of the $E$-field was as follows.
Setting the Wilson coupling to zero $<\partial_ij_i>$ was exactly zero.
This shows that the doublers did not decouple. They
cancel the anomaly and we are left with a vectorlike anomaly free
theory.

Next I included a Wilson coupling $w=0.9$.
To see, whether for the $16^3$ lattice and the choice of $\mu_0$ and $w$
used here, the situation is similar to the case of the $L=100$ system
used in section 2, I calculated the lowest eigenvalue
$\lambda_0$ and the corresponding eigenfunction from the Hamiltonian
(\ref{eq:hamiltonian2}).
I find that the wavefunctions have only a tiny overlap. The support of the
eigenfunctions extends over about 7 $s$-slices.
For small momenta the $16^3$ system still has a chiral fermion and is
appropriate to test the anomaly equation (\ref{eq:latticeanomaly}).

I calculated $<\partial_ij_i>$ from the inverse fermion matrix.
The summation over $s$ (see eq.(\ref{eq:latticedef}))
was taken over three $s$-slices on each side of the $s=1$ slice. Increasing the
number of $s$-slices did not change
the numerical value of $<\partial_ij_i>$. This fits into the picture I
have obtained for the wavefunction from the Hamiltonian.

To test the anomaly equation (\ref{eq:latticeanomaly}) I constructed the
ratio
\begin{equation}
R_{anomaly} =\frac{q^2E_{eff}(t)}{2\pi}/<\partial_ij_i>\; .
\label{eq:eratio}
\end{equation}
In the 3-4-5 model I find find $R_{anomaly}$ to be independent of $t$ for each
fermion with charge $q=3,4,5$, respectively. This shows that the
divergencies of the individual flavor currents are anomalous!
Moreover, if the anomaly equation (\ref{eq:latticeanomaly}) is realized
$R_{anomaly}$ should equal one. I find from the numerical computation of
$<\partial_ij_i>$, $R_{anomaly} \approx 0.98$ in very good agreement with the
theoretical expectations.

In Fig.3 I plot $q<\partial_ij_i>$ for the 3-4-5 model.
The individual $<\partial_ij_i>$ correspond
to the different charges $q$ in the 3-4-5
model as indicated in the figure. They
individually satisfy the relation $R_a =1$ within $2\%$.
Notice that in the case of $q=5$ the sign of $<\partial_ij_i>$ is
reversed showing that the chirality of the zero mode is flipped.
Moreover, the figure also shows the sum of all $<\partial_ij_i>$ as the
straight line at zero! The anomalies cancel,
we end up with a anomaly free gauge current in spite of the fact that the
individual flavor currents are seen to be anomalous.

\section{Conclusion}

In this work I have shown for the first time that it is possible
in practice to see
anomalous behaviour of currents on the lattice. This was possible by
using a recent proposal by Kaplan which circumvents the
Nielson-Ninomyia theorem by starting with a three dimensional vectorlike
theory with a mass defect
and a Wilson term. This mass defect or domain wall
guarantees the existence of a zero mode on the domain wall. Kaplan made use
of this zero mode to construct a chiral gauge theory in the lower
2-dimensional target gauge theory. By introducing a
Wilson-term it is possible to get rid of the unwanted doubler
modes and to end up with only one chiral mode.

By solving the Hamiltonian problem on the finite lattice numerically I
demonstrated that this scenario is true also for the finite lattice
system. I find a chiral fermion for small momenta located at one of the
domain walls and that the lattice system has an energy gap.
I showed by numerically calculating the gauge current on finite
lattices that as a consequence of the existence of the chiral fermion
the anomaly can be seen
on the lattice and that the
divergence of the gauge current $<\partial_ij_i>$,
eq.(\ref{eq:latticecurrent}),
satisfies the anomaly equation eq.(\ref{eq:latticeanomaly}).
In addition I demonstrated
that in the case of an anomaly free theory, which was taken to be the
3-4-5 model, the anomalies on the lattice cancel for the gauge current.

Although these results are certainly only a first step, the results
are promising. They clearly indicate that the
Kaplan proposal for chiral lattice fermions (or regulated chiral
fermions in general) has a chance to solve the old puzzle of chiral
fermions on the lattice.
The next step is to
render the gauge fields dynamical and to see whether one can get out a
2-dimensional
chiral gauge theory -the chiral Schwinger model-
in the continuum limit.
Then one might proceed to four dimensions.
Work in this direction is in progress.

\section*{Acknowledgements}
I want to thank David Kaplan not only for numerous helpful
discussions but also
for sharing and explaining to me his idea of chiral fermions prior to
publication. This work is supported by DOE grant DE-FG-03-90ER40546.
\pagebreak

\pagebreak

\section*{Figure Caption}

{\bf Fig.1} The eigenfunctions corresponding to the lowest eigenvalues
$\pm\lambda_0$ for the smallest lattice momentum,
$k=\sin(\pi/2L)$,  are plotted. The system
size is $L=L_s=100$, the domain wall mas is $\mu_0=0.81$ and the Wilson
coupling $w=0.9$. The wavefunction for $+\lambda_0$ is located at $s=1$,
while the wavefunction for $-\lambda_0$ is at $\frac{L_s}{2}+1$. They show
no overlap.

\noindent {\bf Fig.2a} The two lowest
positive eigenvalues $\lambda_0$ and $\lambda_1$
from the Hamiltonian \protect{(\ref{eq:hamiltonian2})} as a
function of the lattice momenta $k=\sin(\pi(n +\frac{1}{2})/L)$
where $n=0,...,L-1$.
The parameters are as in fig.1. For small $k < k_c\approx 0.9$,
the system has a mass gap,
the lowest eigenvalue exhibits the dispersion
relation of a massless fermion and $\lambda_1$ is at the order of the
cut-off.

\noindent {\bf Fig.2b}
The ratio $R=<\psibar\psi>/<\psibar\sigma_1\psi>$ is shown.
$R$ measures the ``violation of chirality''. It is zero when the
fermion is a chiral eigenstate and the discrepancy from zero indicates
that the state is no longer chiral. The figure clearly
shows that the lowest eigenvalue $\lambda_0$ in fig.2a (which
corresponds also to the wavefunction centered at $s=1$ in fig.1) belongs
to a chiral fermion at low energies.

\noindent {\bf Fig.3}
The divergence of the $1+1$-dimensional gauge current in an
external $E$-field. The system size here is $16^3$. The parameters are
otherwise like in figs.1,2, i.e. $\mu=0.81$ and $w=0.9$. The strength of
the $E$-field is $E_0=0.001$. The different curves correspond to the
contributions to $q<\partial_ij_i>$ ,$i=x,t$,
(see eq.(\ref{eq:latticecurrent})) from fermions of charge
$q=3,4,5$ and chirality $+1,+1,-1$, respectively. The straight line at
zero is the result of the sum of the three individual curves, showing
that the anomalies cancel in the gauge current, even though the
individual flavor currents are anomalous. Note that the results shown in
the figure do not involve a {\em simulation} but stem from a direct numerical
{\em computation} of the gauge current. Accordingly they do not have
errorbars.

\end{document}